\documentclass[preprint,preprintnumbers, prd, floatfix,  superscriptaddress,nofootinbib] {revtex4-1}
\usepackage{graphicx}
\usepackage{epsfig}
\usepackage{bm}
\usepackage{amssymb}
\usepackage{float}
\usepackage{amsmath}
\usepackage{dcolumn}
\usepackage{cancel}
\usepackage[colorlinks]{hyperref}
\usepackage[usenames,dvipsnames]{color}
\usepackage{color}
\usepackage{epstopdf}
\usepackage{nccbbb}

\newcommand{\CD}{{\cal D}}
\newcommand{\CR}{{\cal R}}
\newcommand{\CQ}{{\cal Q}}

\newcommand{\average}[1]{\left\langle #1 \right\rangle_\CD}

\newcommand{\naverage}[1]{\left\langle #1 \right\rangle_{\CD_{\rm \bf 0}}}

\newcommand{\now}[1]{{#1_{\rm \bf 0}}}
\newcommand{\be}{\begin{equation}}
\newcommand{\ee}{\end{equation}}
\newcommand{\bea}{\begin{eqnarray}}
\newcommand{\eea}{\end{eqnarray}}
\newcommand{\bean}{\begin{eqnarray*}}
\newcommand{\eean}{\end{eqnarray*}}
\hypersetup{
     breaklinks=true,
    pdfstartview={FitH},    colorlinks=true,       
    linkcolor=blue,          
    citecolor=red,        
    filecolor=magenta,      
    urlcolor=blue,           
    anchorcolor=green,      
    linktocpage=true
}

\begin{document}
\title{The dark energy phenomenon from backreaction effect }

\author{Yan-Hong Yao}
\email{yhy@mail.nankai.edu.cn}
\author{Xin-He Meng}
\email{xhm@nankai.edu.cn}

\affiliation{Department of Physics, Nankai University, Tianjin 300071, China}

\begin{abstract}
In this paper, we interpret the dark energy phenomenon as an averaged effect caused by small scale inhomogeneities of the universe with the use of the spatial averaged approach of Buchert. Two models are considered here, one of which assumes that the backreaction term ${\cal Q}_\CD$ and the averaged spatial Ricci scalar $\average{\CR}$ obey the scaling laws of the volume scale factor $a_\CD$ at adequately late times, and the other one adopts the ansatz that the backreaction term ${\cal Q}_\CD$ is a constant in the recent universe. Thanks to the effective geometry introduced by Larena et. al. in their previous work, we confront these two backreaction models with latest type Ia supernova and Hubble parameter observations, coming out with the results that the constant backreaction model is slightly favoured over the other model and the best fitting backreaction term in the scaling backreaction model behaves almost like a constant. Also, the numerical results show that the constant backreaction model predicts a smaller expansion rate and decelerated expansion rate than the other model does at redshifts higher than about 1, and both backreaction terms begin to accelerate the universe at a redshift around 0.5.

\end{abstract}

\maketitle

\section{Introduction}
\label{intro}
According to recent observations of type Ia supernovae, the universe is in a state of accelerated expansion \cite{riess1998,perlmutter1999measurements}. The simplest scenario to account for these observations is a positive cosmological constant in Einstein's equations (The most well known cosmology model including such constant is the so called Lambda cold dark matter $(\Lambda CDM) $ model.), which is assumed to be an effect of quantum vacuum fluctuations. However, because of the huge discrepancy between the theoretical expected value and the observed one, other alternative scenarios have been proposed, including scalar field models such as quintessence\cite{Caldwell1998Cosmological}, phantom\cite{Caldwell1999A}, dilatonic\cite{Piazza2004Dilatonic}, tachyon\cite{Padmanabhan2002Accelerated}and quintom\cite{Bo2006Oscillating} etc. and modified gravity models such as braneworlds \cite{maartens2010brane}, scalar-tensor gravity \cite{esposito2001scalar}, higher-order gravitational theories \cite{capozziello2005reconciling,das2006curvature}. Since the so called fitting problem that how well is our universe described by a standard Friedmann-Lema\^{\i}tre-Robertson-Walker(FLRW) model is not solved yet, recently a third alternative has been considered to explain the dark energy phenomenon as an averaged effect caused by small scale inhomogeneities of the universe \cite{rasanen2004dark,kolb2006cosmic}.

In order to consider cosmology model without assuming a FLRW background, it is necessary to answer a longstanding question that how to average a general inhomogeneous model. To date the macroscopic gravity (MG) approach \cite{zalaletdinov1992averaging,zalaletdinov1997averaging,zalaletdinov1993towards,mars1997space} is probably the most well known attempt at averaging in space-time. Although it is the only approach that gives a prescription for the correlation functions which emerge in an averaging of the Einstein's field equations, so far it required a number of assumptions about the correlation functions which make the theory less convictive. Therefore, in this paper we adopt another averaged approach which is put forward by Buchert \cite{buchert2000average,buchert2001average}, in despite of its foliation dependent nature, such approach is quite simple and hence becomes the most well studied theoretical framework of averaged models. Since the averaged field equations in such approach do not form a closed set, one needs to make some assumptions about the backreaction term appeared in the averaged equations. In \cite{buchert2006correspondence}, by taking the assumption that the backreaction term ${\cal Q}_\CD$ and the averaged spatial Ricci scalar $\average{\CR}$ obey the scaling laws of the volume scale factor $a_\CD$, Buchert
proposes a simple backreaction model. To confront such model with observations, Larena et. al. present the effective geometry with the introduction of a template metric that is only compatible with homogeneity and isotropy on large scales of FLRW cosmology instead of on all scales \cite{larena2009testing}. As was pointed out by Larena et. al., the scaling solution cannot be expected to fully represent the realistic backreaction effect throughout the whole history of the universe since we expect that the realistic backreaction term will change considerably at redshift $z_\CD\sim10$. However, since we only use the datasets of type Ia supernova and observational Hubble parameter in this paper, we merely concern the behavior of the backreaction term at adequately late times, i.e. $z_\CD\lesssim O(1)$, which means that although we assume ${\cal Q}_\CD$ obeys scaling laws of $a_\CD$ in such redshift range , it can behave very differently at higher redshifts, particularly, such term encounters rapid change when $z_\CD\sim10$ because of the structure formation effects, and becomes negligible when $z_\CD\gtrsim1000$, which is reasonable because of the consistence between perturbation theory predictions and CMB observations. Nevertheless, we still doubt that the scaling solution is a prime description of the late-time backreaction term, so we propose another parameterization of ${\cal Q}_\CD$ by simply setting it as a constant at late times, and it turn out that such model is preferred by observations. We use the natural units $c=1$ throughout the paper.

The paper is organized as follows. In Section \ref{sec:1}, the spatial averaged approach of Buchert is demonstrated with presentation of the averaged equations for the volume scale factor $a_\CD$. In Section \ref{sec:2}, we introduce the template metric, which is a necessary tool to test the theoretical preditions with observations, and computation of observables. In Section \ref{sec:3}, we apply a simple likelihood analysis of two backreaction models by confronting them with latest type Ia supernova and Hubble parameter observations. After analysis of the results in Section \ref{sec:3}, we summarize our
results in the last section.

\section{The backreaction models}
\label{sec:1}
In \cite{buchert2000average}, Buchert considers a universe filled with irrotational dust with energy density $\varrho$. By foliating space-time with the use of Arnowitt-Deser-Misner(ADM) procedure and defining an averaging operator that acts on any spatial scalar $\Psi$ function as
\begin{equation}
\average{\Psi}:=\frac{1}{V_{\CD}}\int_{\CD}\Psi Jd^{3}X,
\end{equation}
where $V_{\CD}:=\int_{\CD}Jd^{3}X$ is the domain's volume, he obtains two averaged equations here we need, the averaged Raychaudhuri equation
\begin{equation}
3\frac{{\ddot a}_\CD}{a_\CD} + 4\pi G \average{\varrho} ={\CQ}_\CD,
\end{equation}
and the averaged Hamiltonian constraint
\begin{equation}
3\left( \frac{{\dot a}_\CD}{a_\CD}\right)^2 - 8\pi G \average{\varrho}= - \frac{\average{\CR}+{\CQ}_\CD }{2}.
\end{equation}
In these two equations, $a_\CD(t) = \left(\frac{V_\CD (t)}{V_{\now\CD}}\right)^{1/3}$ is the volume scale factor, where $V_{\now\CD} =\vert{\now\CD}\vert$ denotes the present value of the volume, and ${\cal Q}_\CD$, $\average{\CR}$  represent the backreaction term and the averaged spatial Ricci scalar respectively, which are related by the following integrability condition
\begin{equation}
\frac{1}{a_\CD^6}\partial_t \left(\,{\CQ}_\CD \,a_\CD^6 \,\right)
\;+\; \frac{1}{a_\CD^{2}} \;\partial_t \left(\,\average{\CR}a_\CD^2 \,
\right)\,=0\;.
\end{equation}
One can then obtain a specific backreaction model with an extra ansatz about the form of ${\cal Q}_\CD$ and $\average{\CR}$. A popular choice is to assume that
\begin{equation}
{\cal Q}_{\CD}={\cal Q}_{\now\CD}a_{\CD}^{p};
\average{\CR}=\naverage{\CR}a_{\CD}^{n}
\end{equation}
where $n$ and $p$ are real numbers, while ${\cal Q}_{\now\CD}$ and $\naverage{\CR}$ represent the present value of the backreaction term and the averaged spatial Ricci scalar respectively. There are two types of solutions found in \cite{buchert2006correspondence}. The first type, with $n=-2$ and $p=-6$, is less important since at late times it corresponds to a quasi-Friedmannian model in which the backreaction effect can be neglected. The second type, which demands $n=p$, has the
explicit expression as:
\begin{eqnarray}
\average{\CR}&=&\naverage{\CR}a_{\CD}^{n} ,\\
{\cal Q}_{\CD}&=&-\frac{n+2}{n+6}\naverage{\CR}a_{\CD}^{n}.
\end{eqnarray}
As mentioned above, we only assume such parameterization of the backreaction term to be valid in the recent universe.

By introducing the following dimensionless parameters:
\begin{eqnarray}
\Omega_{m}^{\CD} : &=& \frac{8\pi G}{3 H_{\CD}^2} \langle\varrho\rangle_{\cal D} ,\\
\Omega_{\CR}^{\CD} :&=&  - \frac{\average{\CR}}{6 H_{\CD}^2 } ,\\
\Omega_{\CQ}^{\CD} :&=&  - \frac{{\CQ}_{\CD}}{6 H_{\CD}^2 } ,\\
\Omega_X^\CD:&=&\Omega_{\CR}^{\CD} + \Omega_{\CQ}^{\CD} ,
\end{eqnarray}
one can express the volume Hubble parameter $H_\CD : = {\dot a}_\CD / a_\CD$ and the volume deceleration parameter $q^\CD := -\frac{{\ddot a}_\CD}{a_\CD}\frac{1}{H_\now\CD^2}$  as:
\begin{eqnarray}
H_{\CD}(a_{\CD})&=&H_{\now\CD}\left(\Omega_{m}^{\now\CD}a_{\CD}^{-3}+\Omega_{X}^{\now\CD}a_{\CD}^{n}\right)^\frac{1}{2},\\
q^\CD(a_{\CD})&=&\frac{1}{2}\Omega_{m}^{\now\CD}a_{\CD}^{-3}-\frac{1}{2}(n+2)\Omega_{X}^{\now\CD}a_{\CD}^{n}.
\end{eqnarray}
In this paper, we propose another backreaction model with the assumption that the backreaction term is a constant at late times of the universe, which means, by using the integrability condition, ${\cal Q}_\CD$ and $\average{\CR}$ have the following expression:
\begin{eqnarray}
\average{\CR}&=&-3{\cal Q}_{\now\CD}+(3{\cal Q}_{\now\CD}+\naverage{\CR})a_{\CD}^{-2},\\
{\cal Q}_{\CD}&=&{\cal Q}_{\now\CD},
\end{eqnarray}
from which one can obtain the volume Hubble parameter $H_{\CD}$ and the volume deceleration parameter $q^\CD$ in this backreaction model as follow with the use of the averaged equations
\begin{eqnarray}
H_{\CD}(a_{\CD})&=&H_{\now\CD}(\Omega_{m}^{\now\CD}a_{\CD}^{-3}-2\Omega_{\CQ}^{\now\CD}+(\Omega_{\CR}^{\now\CD}+3\Omega_{\CQ}^{\now\CD})a_{\CD}^{-2})^\frac{1}{2},\\
q^\CD(a_{\CD})&=&\frac{1}{2}\Omega_{m}^{\now\CD}a_{\CD}^{-3}+2\Omega_{\CQ}^{\now\CD}.
\end{eqnarray}

\section{Effective geometry}
\label{sec:2}
\subsection{The template metric}
In \cite{larena2009testing}, a template metric was proposed by Larena et. al. as follows,
\begin{equation}
\label{eq:tempmetric1}
{}^4 {\bf g}^\CD = -dt^2 + L_{\now H}^2 \,a_\CD^2 \gamma^\CD_{ij}\,dX^i \otimes dX^j \;\;,
\end{equation}
where $L_{\now H}=1/H_{\now\CD}$ is the present size of the horizon introduced so that the coordinate distance is dimensionless,
and the domain-dependent effective three-metric reads:
\begin{equation}
\gamma^\CD_{ij}\,dX^i \otimes dX^j =\frac{dr^2}{1-\kappa_{\CD}(t)r^2}+r^2d\Omega^{2}
\end{equation}
with $d\Omega^{2}=d\theta^{2}+\sin^{2}(\theta)d\phi^2$, this effective three-template metric is identical to the spatial part of a FLRW metric at any given time, but its scalar curvature  $\kappa_{\CD}$  can vary from time to time. As was pointed out by Larena et. al., $\kappa_{\CD}$ cannot be arbitrary, more precisely, they argue that it should be related to the true averaged scalar curvature $\average{\CR}$ in the way that
\begin{equation}
\average{\CR}=\frac{\kappa_{\CD}(t)|\naverage{\CR}|a_{\now\CD}^{2}}{a_{\CD}^{2}(t)}
\end{equation}
, which is taken as one of the assumptions for two models considered in this paper.

\subsection{Computation of observables}
The computation of effective distances along the light cone defined by the template metric is very different from that of distances in FLRW models. Firstly, let us introduce an effective redshift $z_{\CD}$ defined by
\begin{equation}
1+z_{\CD}:=\frac{(g_{ab}k^{a}u^{b})_{S}}{(g_{ab}k^{a}u^{b})_{O}}\mbox{ ,}
\end{equation}
where the letters O and S denote the evaluation of the quantities at the observer and at the source respectively, $g_{ab}$ in this expression represents the template metric, while $u^{a}$ is the four-velocity of the dust which satisfies $u^{a}u_{a}=-1$, $k^{a}$ the wave vector of a light ray travelling from the source S towards the observer O with the restrictions $k^{a}k_{a}=0$. Then, by normalizing this wave vector such that $(k^{a}u_{a})_{O}=-1$ and introducing the scaled vector $\hat{k}^{a}=a_{\CD}^{2}k^{a}$, we have the following equation:
\begin{equation}
\label{eq:defred2}
1+z_{\CD}=(a_{\CD}^{-1}\hat{k}^{0})_{S}\mbox{ ,}
\end{equation}
with $\hat{k}^{0}$ obeying the null geodesics equation $k^{a}\nabla_{a}k^{b}=0$ which leads to
\begin{equation}
\label{eq:evolk}
\frac{1}{\hat{k}^{0}}\frac{d\hat{k}^{0}}{da_{\CD}}=-\frac{r^{2}(a_{\CD})}{2(1-\kappa_{\CD}(a_{\CD})r^{2}(a_{\CD}))}\frac{d\kappa_{\CD}(a_{\CD})}{da_{\CD}}\mbox{ .}
\end{equation}

As usual, the coordinate distance can be derived from the equation of radial null geodesics:
\begin{equation}
\label{eq:coorddist}
\frac{dr}{da_{\CD}}=-\frac{H_{\now\CD}}{a_{\CD}^{2}H_{\CD}(a_{\CD})}\sqrt{1-\kappa_{\CD}(a_{\CD})r^{2}}
\end{equation}

Solving these two equations with the initial condition $\hat{k}^{0}(1)=1, r(1)=0 $ and then plugging $\hat{k}^{0}(a_\CD)$ into Eq.~(\ref{eq:defred2}), one finds the relation between the redshift and the scale factor. With these results, we can determine the volume Hubble parameter $H_{\CD}(z_{\CD})$ and the luminosity distance $d_{L}(z_{\CD})$ of the sources defined by the following formula
\begin{eqnarray}
\label{eq:distances}
d_{L}(z_{\CD})&=&\frac{1}{H_{\now\CD}}(1+z_{\CD})^{2}a_{\CD}(z_{\CD})r(z_{\CD}).
\end{eqnarray}
Having computed these two observables , it is then possible to compare the backreaction model predictions with type Ia supernova and Hubble parameter observations.

\section{Constraints from supernovae data and OHD}
\label{sec:3}
In this section, we perform a simple likelihood analysis on the free parameters of two backreaction model mentioned above with the combination of datasets from type Ia supernova and Hubble parameter observations.

The best-fit values of the model parameters $(\Omega_{m}^{\now\CD},o)$ (Here $o$ represents $n$ in the case of scaling backreaction model and $\Omega_{\CR}^{\now\CD}$ in the case of constant backreaction model respectively.) from the recently released Union2.1\cite{Suzuki2012The} compilation with 580 data points are determined by minimizing
\begin{equation}\label{}
  \chi_{SN Ia}^{2}(\Omega_{m}^{\now\CD},o)= R-\frac{S^{2}}{T}
\end{equation}
here $R$, $S$ and $T$ are defined as
\begin{eqnarray}
  R &=& \sum_{i=0}^{580}\frac{\left(5\log_{10}\left[H_{\now\CD}d_{L}({z_{\CD}}_{i})\right]-\mu_{obs}({z_{\CD}}_{i})\right)^{2}}{\sigma_{\mu}^{2}({z_{\CD}}_{i})},\\
  S &=& \sum_{i=0}^{580}\frac{\left(5\log_{10}\left[H_{\now\CD}d_{L}({z_{\CD}}_{i})\right]-\mu_{obs}({z_{\CD}}_{i})\right)}{\sigma_{\mu}^{2}({z_{\CD}}_{i})}, \\
  T &=& \sum_{i=0}^{580}\frac{1}{\sigma_{\mu}^{2}({z_{\CD}}_{i})}.
\end{eqnarray}

where $\mu_{obs}$ represents the observed distance modulus and $\sigma_{\mu}$ denotes its statistical uncertainty.

For the observed Hubble parameter dataset in Table \ref{tab:1}, the best-fit values of the parameters $(H_{\now\CD},\Omega_{m}^{\now\CD},o)$ can be determined by a likelihood analysis based on the calculation of
\begin{equation}
 \chi^{2}_H(H_{\now\CD},\Omega_{m}^{\now\CD},o)=\sum_{i=0}^{30} \frac{(H_{\CD}({z_{\CD}}_{i};H_{\now\CD},\Omega_{m}^{\now\CD},o)-H_{obs}({z_{\CD}}_{i}))^2}{\sigma_{H}^{2}({z_{\CD}}_{i})}.
\end{equation}
As Ma et. al.\cite{ma2011power} stated, the marginalized probability density function determined by integrating $ \rm e^{-\frac{\chi^{2}_H(H_{\now\CD},\Omega_{m}^{\now\CD},o)}{2}} $ over $H_{\now\CD}$ from $x$ to $y$  with a uniform prior reads
\begin{equation}
  {\rm e^{-\frac{\chi^{2}_H(\Omega_{m}^{\now\CD},o)}{2}}} = \frac{U(x,C,D)-U(y,C,D)}{\sqrt{C}}{\rm e}^{\frac{D^2}{C}}
\end{equation}
where
\begin{equation}
  C=\sum_{i=0}^{30}\frac{H_{\CD}^2({z_{\CD}}_{i};H_{\now\CD},\Omega_{m}^{\now\CD},o)}{2H_{\now\CD}^2\sigma_{H}^{2}({z_{\CD}}_{i})}, D=\sum_{i=0}^{30}\frac{H_{\CD}({z_{\CD}}_{i};H_{\now\CD},\Omega_{m}^{\now\CD},o)H_{obs}({z_{\CD}}_{i})}{2H_{\now\CD}\sigma_{H}^{2}({z_{\CD}}_{i})},
\end{equation}
and
\begin{equation*}
  U(x,\alpha,\beta)={\rm erf}(\frac{\beta-x\alpha}{\sqrt{\alpha}}),
\end{equation*}
$[x,y]$ is taken as $[50,90]$, and erf represents the error function.

Finally, the total $\chi^{2}(\Omega_{m}^{\now\CD},o)$ for the combined observational dataset are given by $\chi^{2}(\Omega_{m}^{\now\CD},o)=\chi_{SN Ia}^{2}(\Omega_{m}^{\now\CD},o)+\chi_{H}^{2}(\Omega_{m}^{\now\CD},o)$.

\begin{figure}
\begin{flushleft}
\begin{minipage}{0.45\linewidth}
  \centerline{\includegraphics[width=1\textwidth]{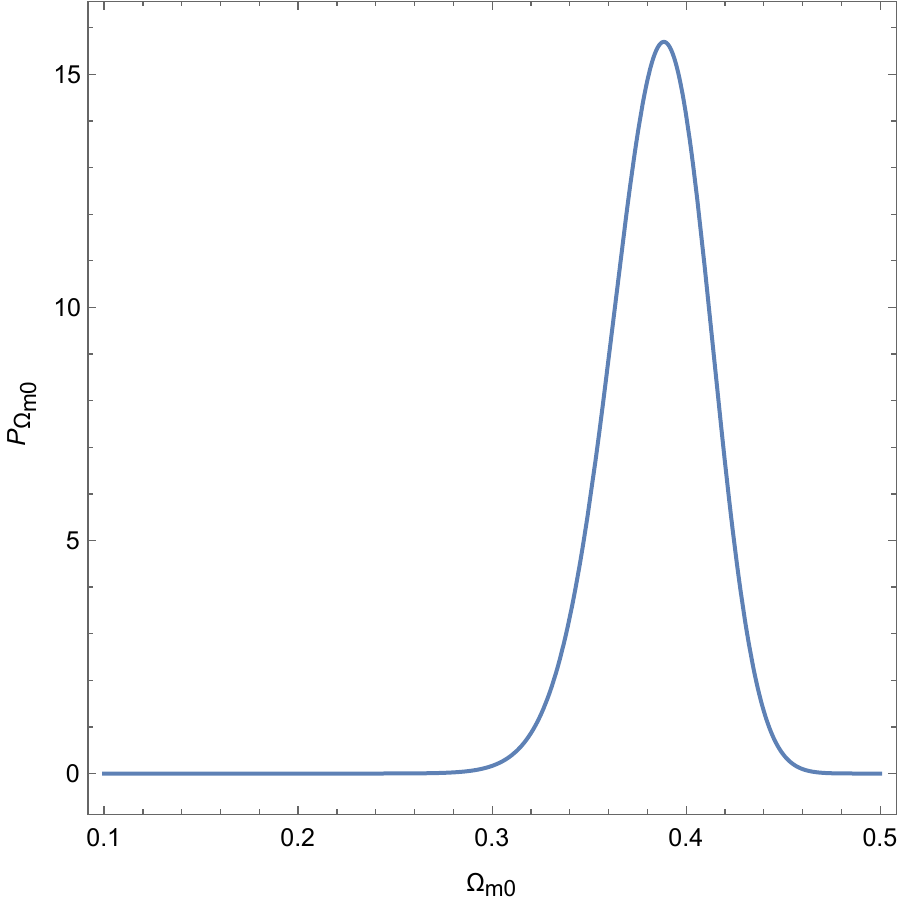}}
\end{minipage}
\hfill
\end{flushleft}

\begin{flushleft}
\begin{minipage}{0.46\linewidth}
  \centerline{\includegraphics[width=1\textwidth]{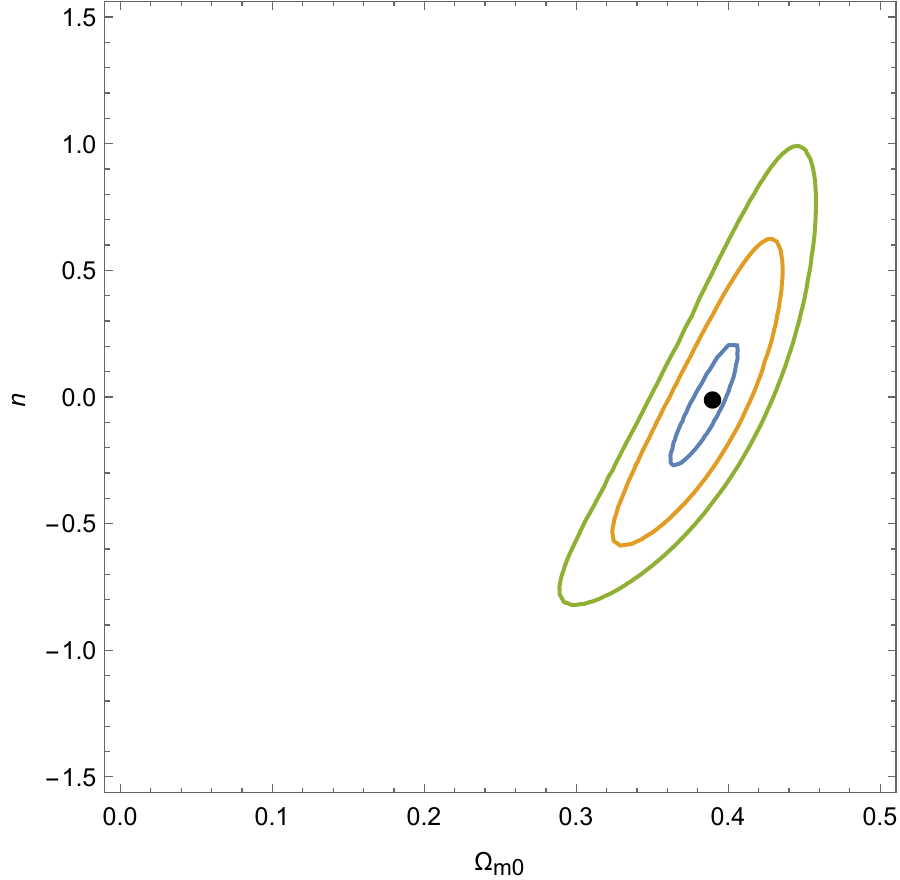}}
\end{minipage}
\hfill
\begin{minipage}{0.45\linewidth}
 \centerline{\includegraphics[width=1\textwidth]{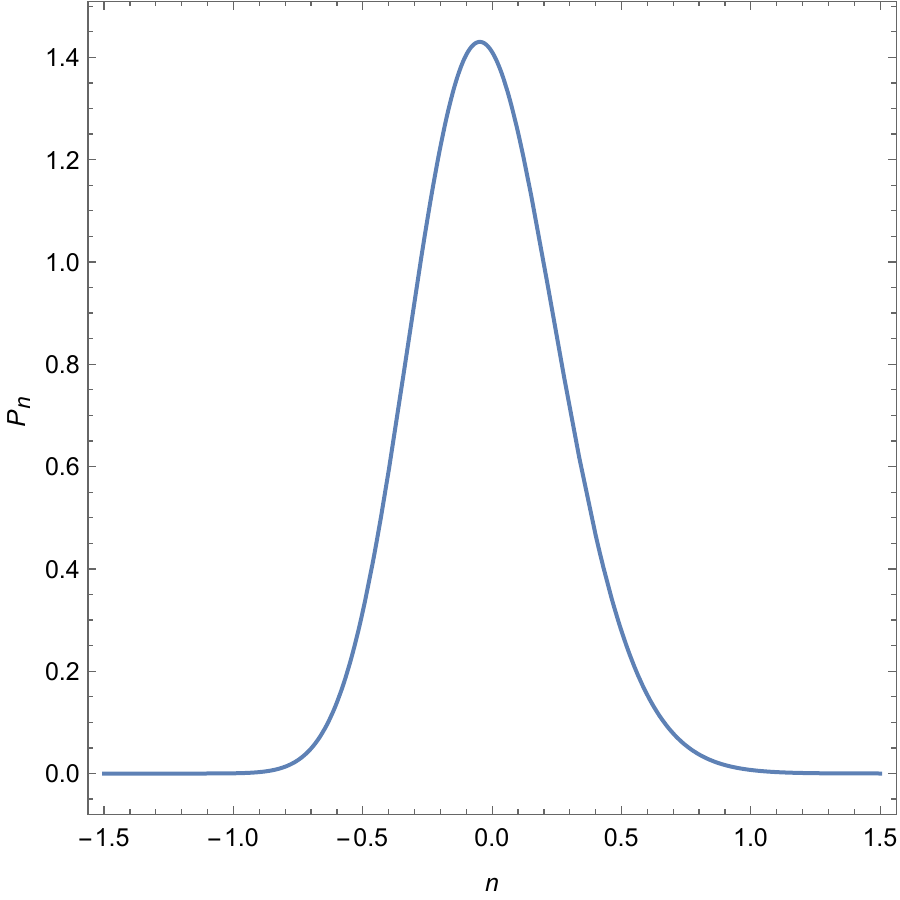}}
\end{minipage}
\end{flushleft}

\caption{The $1\sigma$, $2\sigma$ and $3\sigma$ confidence regions and best fitting point of of the free parameters $\Omega_{m}^{\now\CD},n$ for the
 scaling backreaction model, along with their own probability density function.The prior for $\Omega_{m}^{\now\CD}\varpropto $ H(0.5-$\Omega_{m}^{\now\CD}$)H($\Omega_{m}^{\now\CD}$-0.1),
The prior for $n\varpropto$ H(1.5-$n$)H($n$-(-1.5)),
H(x)is the step function}
\label{fig:1}
\end{figure}

\begin{figure}
\begin{flushleft}
\begin{minipage}{0.45\linewidth}
  \centerline{\includegraphics[width=1\textwidth]{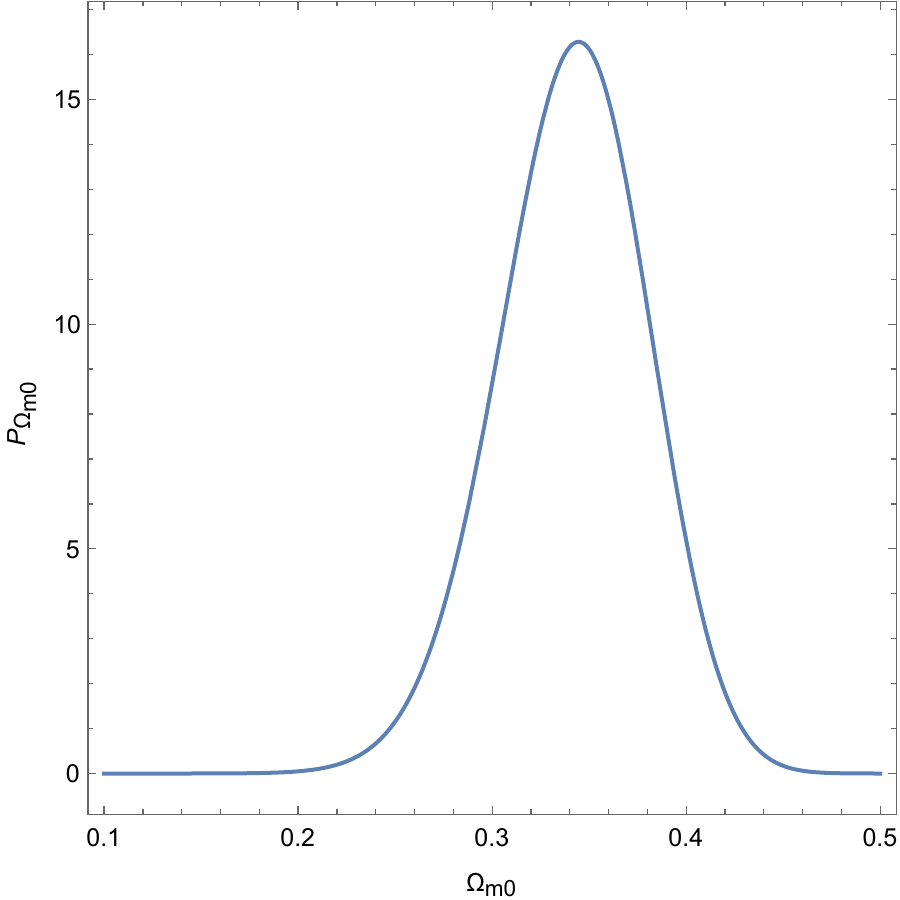}}
\end{minipage}
\hfill
\end{flushleft}

\begin{flushleft}
\begin{minipage}{0.46\linewidth}
  \centerline{\includegraphics[width=1\textwidth]{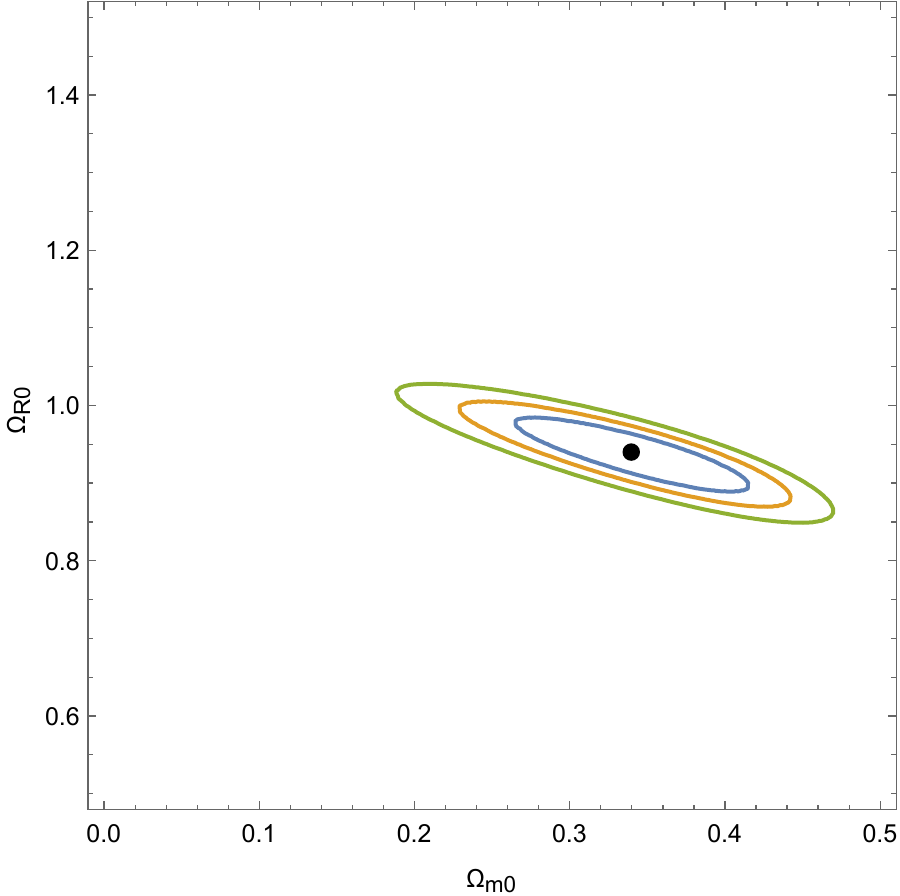}}
\end{minipage}
\hfill
\begin{minipage}{0.45\linewidth}
 \centerline{\includegraphics[width=1\textwidth]{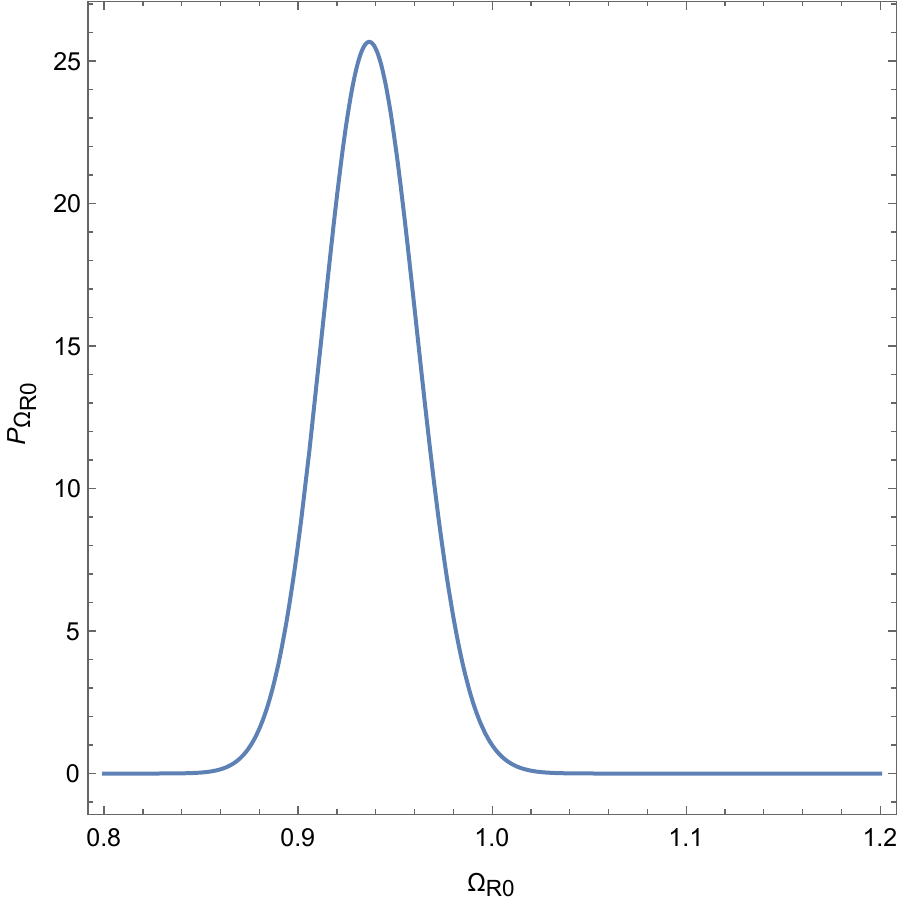}}
\end{minipage}
\end{flushleft}

\caption{The $1\sigma$, $2\sigma$ and $3\sigma$ confidence regions and best fitting point of the free parameters $\Omega_{m}^{\now\CD},\Omega_{\CR}^{\now\CD}$ for the
 constant backreaction model, along with their own probability density function.The prior for $\Omega_{m}^{\now\CD}\varpropto $ H(0.5-$\Omega_{m}^{\now\CD}$)H($\Omega_{m}^{\now\CD}$-0.1),
The prior for $\Omega_{\CR}^{\now\CD}\varpropto$ H(1.2-$\Omega_{\CR}^{\now\CD}$)H($\Omega_{\CR}^{\now\CD}$-0.8),
H(x)is the step function}
\label{fig:2}
\end{figure}

The fitting results attained from analyzing $\chi^{2}(\Omega_{m}^{\now\CD},o)$ by using functions Findminimum and ContourPlot in mathematica are presented in Fig.\ref{fig:1}, Table \ref{tab:2} for the scaling backreaction model and Fig.\ref{fig:2}, Table \ref{tab:3} for the constant backreaction model. The comparison of $\chi^{2}_{min}$ in these two tables show that the constant backreaction model is slightly favoured over the other model by current observations, confirming the correctness of our speculation that the scaling solution is not a prime description of the late-time backreaction term. Also, the result in Table \ref{tab:2} suggests that, the best fitting backreaction term in the scaling backreaction model behaves almost like a constant, which demonstrates the rationality to propose the constant backreaction model rather than other backreaction model in the beginning. Unlike the scaling backreaction model, according to Table \ref{tab:3} and (14), observations favor a monotone decrease averaged spatial Ricci scalar in the constant backreaction model, this dynamical behavior of the averaged spatial Ricci scalar reduces the best fitting value of the model parameter $\Omega_{m}^{\now\CD}$.

\begin{figure}
\begin{minipage}{0.45\linewidth}
  \centerline{\includegraphics[width=1\textwidth]{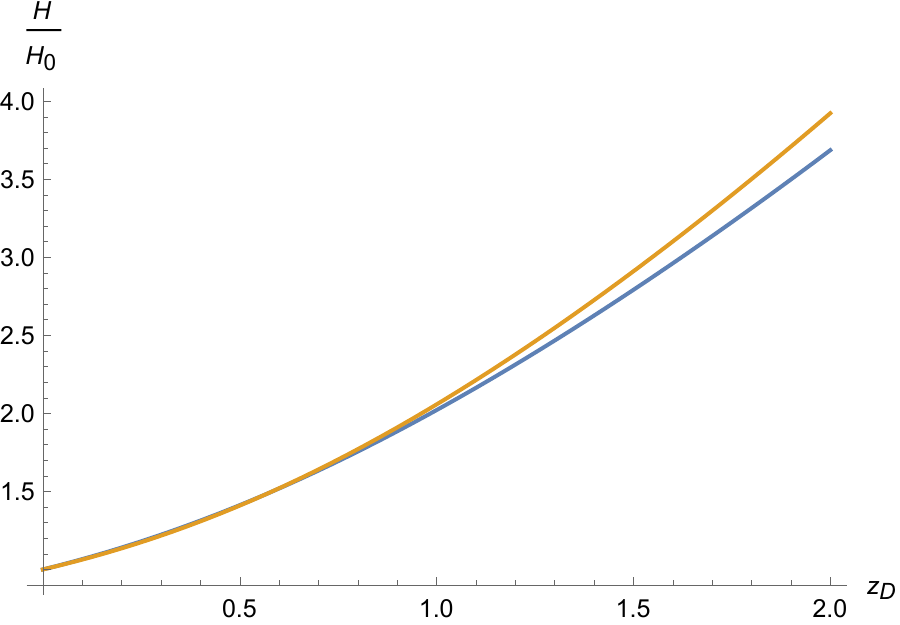}}
\end{minipage}
\hfill
\begin{minipage}{0.45\linewidth}
  \centerline{\includegraphics[width=1\textwidth]{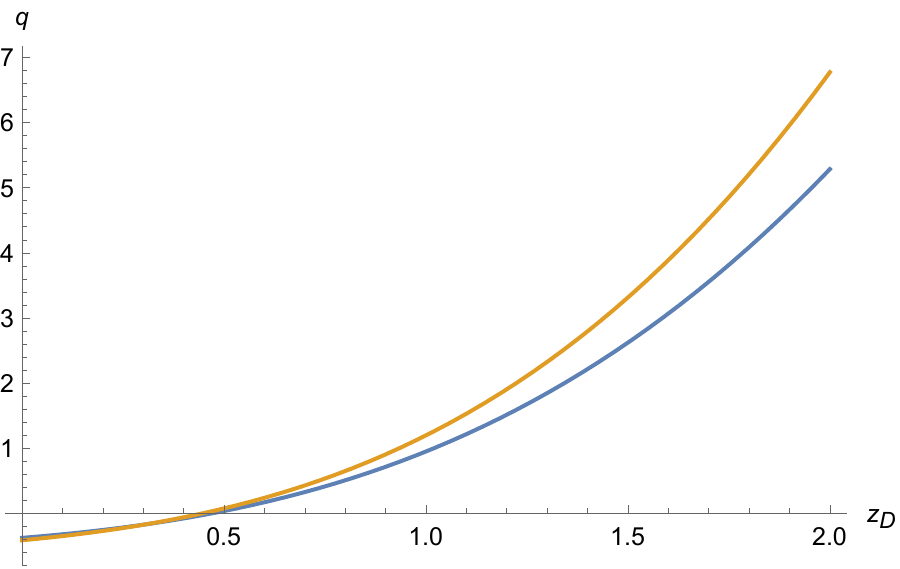}}
\end{minipage}

\caption{The evolution of $\frac{H_{\CD}}{H_{\now\CD}}$  and  $q_{\CD}$ with respect to $z_{\CD}$. Here the blue line and the orange line corresponding to that of the scaling backraction model and the constant backreaction model with best-fit parameters. }
\label{fig:3}
\end{figure}

Noting from the fitting results that the best-fit value of the matter density parameter in the scaling backreaction model is bigger than that in the other model, indicating that this model predicts a larger expansion rate and decelerated expansion rate at high redshifts. Such departure is shown in Fig.\ref{fig:3}, which also reveals that the universes described by two models with their best-fit parameters share the almost same expansion rate and decelerated expansion rate(accelerated expansion rate) once $z_{\CD}$ drops below about 1, and enter a stage of an accelerated expansion with a redshift around 0.5.

\begin{table}
\centering
\begin{tabular}{|lcc|}
\hline
{$z$}   & $H(z)$ &  Ref.\\
\hline
$0.0708$   &  $69.0\pm19.68$         &  Zhang et al. (2014)-\cite{Zhang2014}   \\
    $0.09$       &  $69.0\pm12.0$            &  Jimenez et al. (2003)-\cite{Jimenez2003}   \\
    $0.12$       &  $68.6\pm26.2$           &  Zhang et al. (2014)-\cite{Zhang2014}  \\
    $0.17$       &  $83.0\pm8.0$             &  Simon et al. (2005)-\cite{Simon2005}     \\
    $0.179$     &  $75.0\pm4.0$           &  Moresco et al. (2012)-\cite{Moresco2012}     \\
    $0.199$     &  $75.0\pm5.0$            &  Moresco et al. (2012)-\cite{Moresco2012}     \\
    $0.20$         &  $72.9\pm29.6$         &  Zhang et al. (2014)-\cite{Zhang2014}   \\
    $0.27$       &  $77.0\pm14.0$         &    Simon et al. (2005)-\cite{Simon2005}   \\
    $0.28$       &  $88.8\pm36.6$        &  Zhang et al. (2014)-\cite{Zhang2014}   \\
    $0.352$     &  $83.0\pm14.0$          &  Moresco et al. (2012)-\cite{Moresco2012}   \\
    $0.3802$     &  $83.0\pm13.5$         &  Moresco et al. (2016)-\cite{Moresco2016}   \\
    $0.4$         &  $95\pm17.0$             &  Simon et al. (2005)-\cite{Simon2005}     \\
    $0.4004$     &  $77.0\pm10.2$          &  Moresco et al. (2016)-\cite{Moresco2016}   \\
    $0.4247$     &  $87.1\pm11.2$         &  Moresco et al. (2016)-\cite{Moresco2016}   \\
    $0.4497$     &  $92.8\pm12.9$        &  Moresco et al. (2016)-\cite{Moresco2016}   \\
    $0.4783$     &  $80.9\pm9.0$         &  Moresco et al. (2016)-\cite{Moresco2016}   \\
    $0.48$       &  $97.0\pm62.0$         &  Stern et al. (2010)-\cite{Stern2010}     \\
    $0.593$     &  $104.0\pm13.0$        &  Moresco et al. (2012)-\cite{Moresco2012}   \\
    $0.68$       &  $92.0\pm8.0$        &  Moresco et al. (2012)-\cite{Moresco2012}   \\
    $0.875$     &  $125.0\pm17.0$       &  Moresco et al. (2012)-\cite{Moresco2012}   \\
    $0.88$       &  $90.0\pm40.0$         &  Stern et al. (2010)-\cite{Stern2010}     \\
    $0.9$         &  $117.0\pm23.0$        &  Simon et al. (2005)-\cite{Simon2005}  \\
    $1.037$     &  $154.0\pm20.0$          &  Moresco et al. (2012)-\cite{Moresco2012}   \\
    $1.3$         &  $168.0\pm17.0$        &  Simon et al. (2005)-\cite{Simon2005}     \\
    $1.363$     &  $160.0\pm33.6$          &  Moresco (2015)-\cite{Moresco2015}  \\
    $1.43$       &  $177.0\pm18.0$         &  Simon et al. (2005)-\cite{Simon2005}     \\
    $1.53$       &  $140.0\pm14.0$        &  Simon et al. (2005)-\cite{Simon2005}     \\
    $1.75$       &  $202.0\pm40.0$         &  Simon et al. (2005)-\cite{Simon2005}     \\
    $1.965$     &  $186.5\pm50.4$         &   Moresco (2015)-\cite{Moresco2015}  \\
\hline
\end{tabular}
\caption{\label{tab:1} The current available OHD dataset.}
\end{table}

\begin{table}
\begin{center}
\begin{tabular}{cc|  cc }
\hline\hline scaling backreaction model & &  1$\sigma$ confidence interval
\\ \hline
$\Omega_{m}^{\now\CD}$    && $ 0.39_{-0.03}^{+0.02}$
                     \\
$n$         &&  $-0.01_{-0.31}^{+0.27}$
                     \\
 \hline $\chi^{2}_{min}$  &&  $-851.50$
                      \\
\hline\hline
\end{tabular}
\caption{\label{tab:2} The fitting results of the parameters ($\Omega_{m}^{\now\CD},n$) with 1$\sigma$ region in the scaling backreaction model, $\chi^{2}_{min}$ is corresponding to $(\Omega_{m}^{\now\CD},n)=(0.39,-0.01)$.}
\end{center}
\end{table}

\begin{table}
\begin{center}
\begin{tabular}{cc|  cc }
\hline\hline constant backreaction model & &  1$\sigma$ confidence interval
\\ \hline
$\Omega_{m}^{\now\CD}$    &&  $0.33_{-0.03}^{+0.05}$
                     \\
$\Omega_{\CR}^{\now\CD}$         &&  $0.93_{-0.03}^{+0.03}$
                     \\
 \hline $\chi^{2}_{min}$  &&  $-853.07$
                      \\
\hline\hline
\end{tabular}
\caption{\label{tab:3} The fitting results of the parameters ($\Omega_{m}^{\now\CD},\Omega_{\CR}^{\now\CD}$) with 1$\sigma$ region in the constant backreaction model, $\chi^{2}_{min}$ is corresponding to $(\Omega_{m}^{\now\CD},\Omega_{\CR}^{\now\CD})=(0.33,0.93)$.}
\end{center}
\end{table}

\section{Conclusion and discussion}
\label{sec:4}
In this paper, the dark energy phenomenon has been interpreted as an averaged effect caused by small scale inhomogeneities of the universe. In order to understand the averaged evolutional behavior of the universe within the approach of Buchert, we have considered two backreaction models, one of which assumes that the backreaction term ${\cal Q}_\CD$ and the averaged spatial Ricci scalar $\average{\CR}$ obey the scaling laws of the volume scale factor  $a_\CD$ at adequately late times, and the other one adopts the ansatz that ${\cal Q}_\CD$ is a constant in the recent universe. With the aid of the effective geometry introduced by Larena et. al. in their previous work, we have confronted these two backreaction models with latest type Ia supernova and Hubble parameter observations, and found that the best fitting  backreaction term in the scaling backreaction model behaves almost like a constant, which demonstrates the rationality to propose the constant backreaction model rather than other backreaction model in this paper. Moreover, as is shown by the results of numerical analysis, the constant backreaction model predicts a smaller expansion rate and decelerated expansion rate than the other model does at redshifts higher than about 1 and both backreaction terms begin to accelerate the universe at a redshift around 0.5.

Although we only make assumptions about the specific form of the backreaction term at late times throughout the paper, a complete backreaction model must consider the specific behavior of the backreaction term at arbitrary redshift. Nevertheless, parameterization of the late-time backreaction term is helpful and necessary for searching a complete backreaction model that is also favoured by observations at high redshifts.

\section*{Acknowledgments}
The paper is partially supported by the Natural Science Foundation of China.

\bibliographystyle{spphys}
\bibliography{backreaction}

\end{document}